\documentclass[CEJP,PDF]{cej} 
\usepackage{layout}
\usepackage{amsmath}
\usepackage{textcomp}
\usepackage{hyperref}
\usepackage{graphicx}
\usepackage{epsfig}

\title{Bulk viscosity in heavy ion collision}

\articletype{Research Article}

\author{Victor~Roy \email{victor@vecc.gov.in} and
        A.K.~Chaudhuri
        }
\institute{
     Variable Energy Cyclotron Centre,\\
     1/AF, Bidhannagar,Kolkata-700064 , India
          }

\abstract{The effect of a temperature dependent bulk viscosity to 
entropy density ratio~($\zeta/s$) along with a constant shear viscosity to entropy 
density ratio~($\eta/s$) on the space time evolution of the fluid produced in high
energy heavy ion collisions have been studied in a relativistic viscous hydrodynamics model. 
The boost invariant Israel-Stewart theory of causal relativistic viscous hydrodynamics 
is used to simulate the evolution of the fluid in 2 spatial and 1 temporal dimension.
The dissipative correction to the freezeout distribution for bulk viscosity is calculated 
using Grad's fourteen moment method. From our simulation we show that the method is applicable 
only for $\zeta/s<0.004$. }

\keywords{Bulk Viscosity \*\ Relativistic hydrodynamics \*\ Grad's moment}
\pacs{12.38.Mh  ,47.75.+f,  25.75.Ld}

\begin{document}
\maketitle

\section{Introduction}
Recent experiments in high energy nuclear 
collisions at relativistic
heavy ion collider~(RHIC) confirms the existence of a new state
of matter known as Quark Gluon Plasma~(QGP)~\cite{journal-1}.
The production of QGP in heavy ion collision and its subsequent collective evolution
provide us the unique opportunity to study the transport properties of this most fundamental
form of matter. Relativistic viscous hydrodynamics simulations of observables like elliptic flow 
($v_{2}$) and transverse momentum ($p_{T}$) spectra have been compared to experimental data to
extract the QGP $\eta/s$. Most studies show that the estimated value of $\eta/s$ lies between
$1-4\times(1/4\pi)$. However to correctly extract the $\eta/s$ of the QGP fluid, it is important 
to know the effect of finite bulk viscosity on the fluid evolution. Theoretical 
calculations based on pQCD~\cite{journal-2} and lattice~QCD~\cite{journal-3} 
shows that the bulk viscosity is non-zero for the temperature range applicable 
in the heavy ion collision.  
In this work we use a temperature dependent form of $\zeta/s$ to study the
effect of bulk viscosity in fluid evolution. The dissipative correction to 
the freezeout distribution function bulk viscosity 
has also been considered using Grad's~14 moment method.

\begin{figure}
\includegraphics[width=0.3\textwidth]{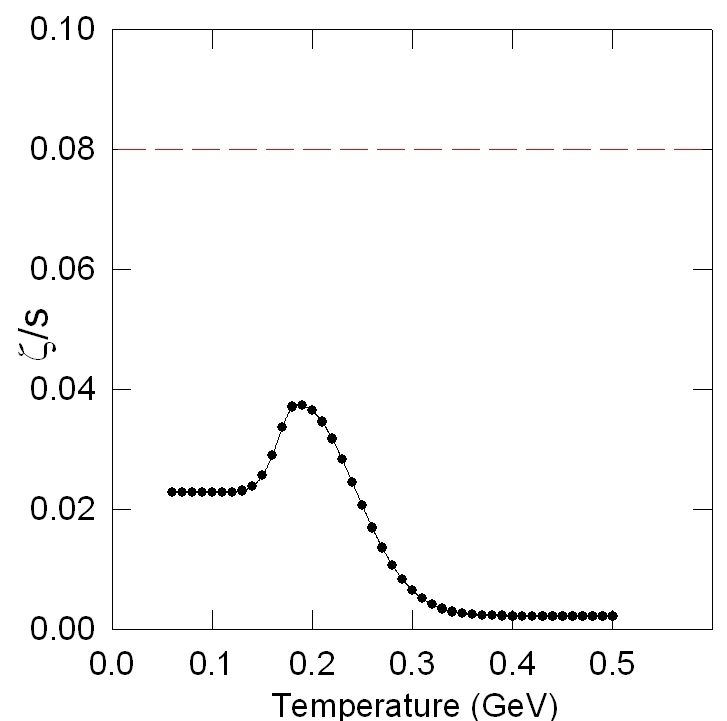}
\caption{ (Color online) $\zeta/s$ as a function of temperature.
Red dashed line is the $\eta/s=1/4\pi$. 
\label{fig1}}
\end{figure}


\section{Viscous hydrodynamic model}
The space time evolution of the fluid was simulated by simultaneously solving 
the energy momentum conservation equation $\partial_{\mu}T^{\mu\nu}=0$,
along with the relaxation equation for shear and bulk stress. Here  
$T^{\mu\nu}=(\epsilon+p+\Pi)u^{\mu}u^{\nu}-(p+\Pi)g^{\mu\nu}+\pi^{\mu\nu}$ is 
the energy momentum tensor and $\epsilon,p$  are energy density, pressure
of the fluid,$g^{\mu\nu}$ is the metric tensor; $\Pi$ and $\pi^{\mu\nu}$ are bulk
and shear stress tensor respectively. According to the Israel-Stewart theory of causal
viscous hydrodynamics~\cite{journal-4}, the shear and bulk viscosity obey 
the following relaxation equations
$D\pi^{\mu\nu}=\frac{1}{\tau_\pi}[2\eta\nabla^{<^{\mu}u^{\nu}>}-\pi^{\mu\nu}]-
\left(u^{\mu}\pi^{\nu\lambda}+u^{\nu}\pi^{\mu\lambda}\right)Du_{\lambda}$; and
$D\Pi=-\frac{1}{\tau_{\Pi}}[\Pi+\zeta\nabla_{\mu}u^{\mu}+\frac{1}{2}\zeta T\Pi\partial_{\mu}(\frac{\tau_{\Pi}u^{\mu}}{\zeta T})]$. Here D is the convective 
derivative,$\tau_{\pi}$ and $\tau_{\Pi}$ are the relaxation time for shear and bulk
stresses respectively. We assume that the fluid achieve near local thermalization at proper time 
0.6 fm. Initial transverse velocity~($v_{T}$) is assumed to be zero.
The initial energy density profile in transverse plane is calculated from a two component Glauber model
with a central energy density $\epsilon_{0}=30 GeV/fm^{3}$. Initial value of $\pi^{\mu\nu}$
and $\Pi$ was set to their corresponding Navier-Stokes estimate. We 
assume the fluid freezes out when an element of it cools down below a constant 
temperature $T_{fo}=130$ MeV. The freezeout procedure was carried out by using Cooper-Frey
algorithm. In the present study, we have used an equation of state~(EoS) where the Wuppertal-Budapest 
lattice calculation~\cite{journal-3} for the deconfined phase is smoothly joined at 
crossover temperature 174 MeV, with hadronic resonance gas EoS comprising all the 
resonances below mass $m_{res}$=2.5 GeV. $\zeta/s$ and $\eta/s$ are inputs to 
viscous hydrodynamics simulation. Figure~\ref{fig1} shows the $\zeta/s$(T), where
$\zeta/s$ in the QGP phase is obtained by using pQCD formula
$\zeta/s=15\frac{\eta}{s}(T)(1/3-c^{2}_{s}(T))^{2}$,
the squared speed of sound $c^{2}_{s}$ was calculated from lattice data~\cite{journal-5}.
In the hadronic phase $\zeta/s$ is parametrized from~\cite{journal-6}.
The red dashed line in figure \ref{fig1} is $\eta/s$.

\section{Results and discussion}
\begin{figure}
\includegraphics[width=0.47\textwidth]{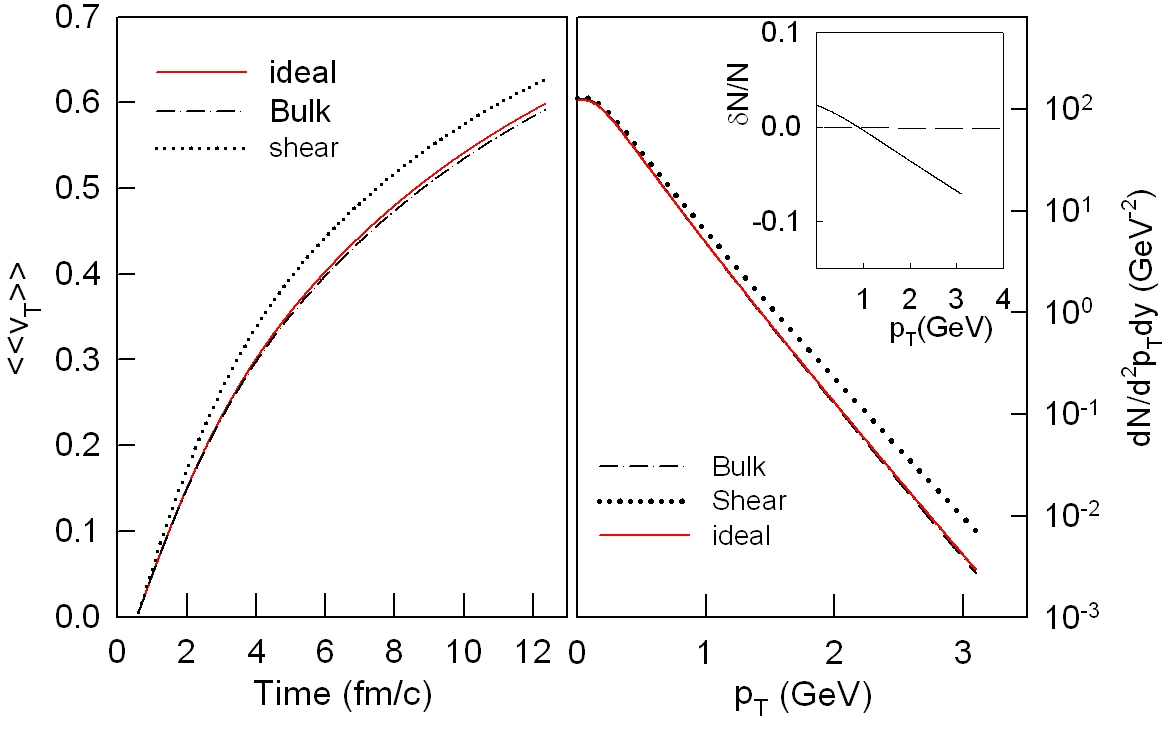}
\caption{(Color online) The left plot shows the temporal evolution of fluid transverse velocity
for ideal~(red),bulk~(dashed dot) and shear viscous~(dotted) evolution. The right plot
is the $\pi^{-}$ invariant yield as a function of $p_{T}$ for ideal~(red)
,bulk~{dashed dot} and shear viscous~(dotted) evolution. The inset figure shows the relative
correction to invariant yield due to the bulk viscosity in comparison to ideal fluid.
\label{fig2}}
\end{figure}

We first discuss the change in pion $p_{T}$ spectra and $v_{2}$ due 
to bulk and shear viscosity in the fluid evolution only.  
In the left panel of figure~\ref{fig2} temporal evolution of spatially averaged 
transverse velocity $\left\langle \left\langle v_{T}\right\rangle\right\rangle
=\frac{\left\langle \left\langle \gamma\sqrt{v^{2}_{x}+v^{2}_{y}}~
\right\rangle\right\rangle}{\left\langle \left\langle \gamma\right\rangle\right\rangle}$
is shown for ideal, shear and bulk
viscous fluid. Here the angular bracket denotes space average and 
$\gamma=\frac{1}{\sqrt{1-v^{2}}}$.
Because of the reduced pressure in bulk viscous evolution,
$\left\langle \left\langle v_{T}\right\rangle\right\rangle$ is reduced 
in comparison to ideal fluid evolution.
Whereas shear viscosity increase the pressure in the transverse direction,
as a result the $\left\langle \left\langle v_{T}\right\rangle\right\rangle$
is larger compared to ideal fluid. The effect of 
the changed fluid velocity in viscous evolution is reflected in the slope 
of the $p_{T}$ spectra of $\pi^{-}$ shown in the right panel of  
figure~\ref{fig2}. The relative change
in the $\pi^{-}$ invariant yield $(\delta N/N_{ideal},\delta N=N_{bulk}-N_{ideal})$
due to the bulk viscosity in comparison to ideal 
fluid is shown in the inset of right plot of figure~\ref{fig2}.
The relative correction is within $\sim10\%$.

\begin{figure}
\includegraphics[width=0.47\textwidth]{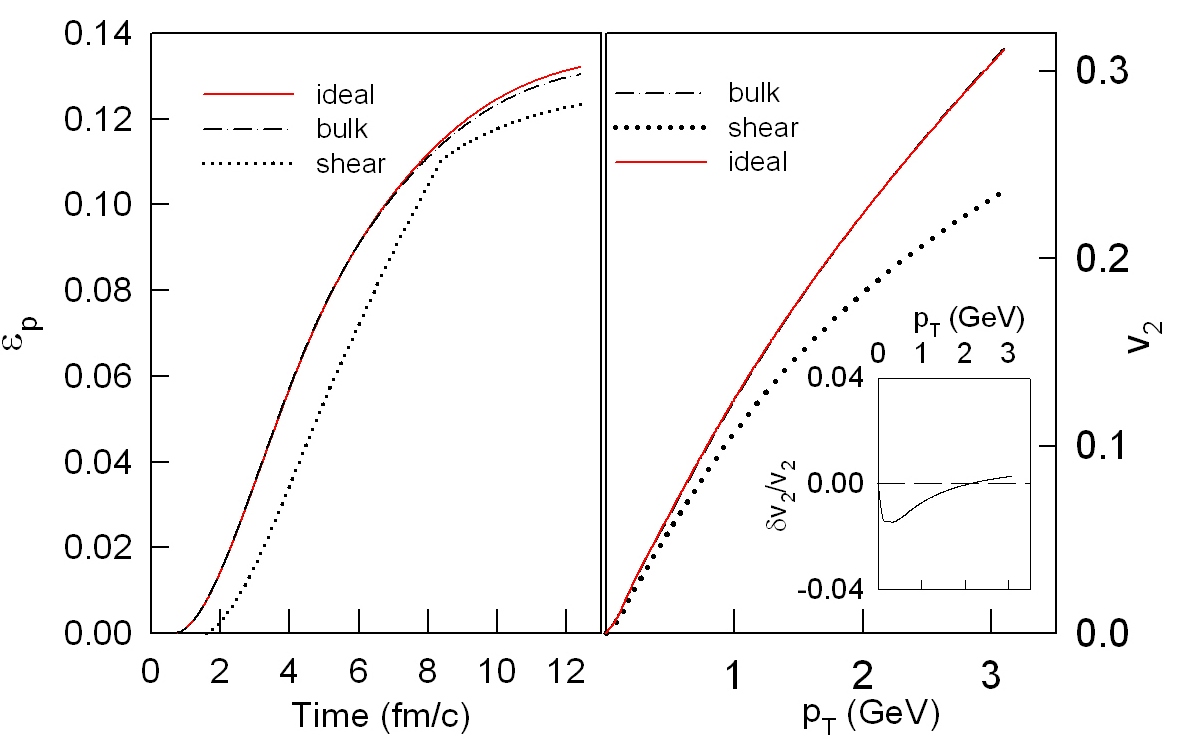}
\caption{(Color online) The left plot is the temporal evolution of momentum anisotropy of the fluid
for ideal(red),bulk(dashed dot) and shear viscous(dotted) evolution. Right plot is the
corresponding elliptic flow~($v_{2}$) of $\pi^{-}$. The inset plot
shows the relative change in $v_{2}$ due to bulk viscosity compared to ideal fluid.\label{fig3}}
\end{figure}

The temporal evolution of momentum space anisotropy $\epsilon_{p}=\frac{\int dxdy (T^{xx}-T^{yy})}
{\int dxdy (T^{xx}+T^{yy})}$ is shown in the left panel of figure~\ref{fig3}.
Viscosity tries to diminish any velocity gradient present in the fluid, as a result
of that $\epsilon_{p}$ is smaller for both shear and 
bulk viscous evolution compared to ideal fluid. In a hydrodynamic model $v_{2}$ is
proportional to $\epsilon_{p}$ hence a reduction in $\epsilon_{p}$ will result in a reduced
$v_{2}$. Elliptic flow of $\pi^{-}$
as a function of $p_{T}$ is shown for ideal, shear and bulk viscous evolution in 
the right plot of figure~\ref{fig3}.
The inset shows the relative correction to $v_{2}$ due to bulk viscosity.
The relative correction to $v_{2}$ is within $\sim3\%$.

\begin{figure}
\includegraphics[width=0.45\textwidth]{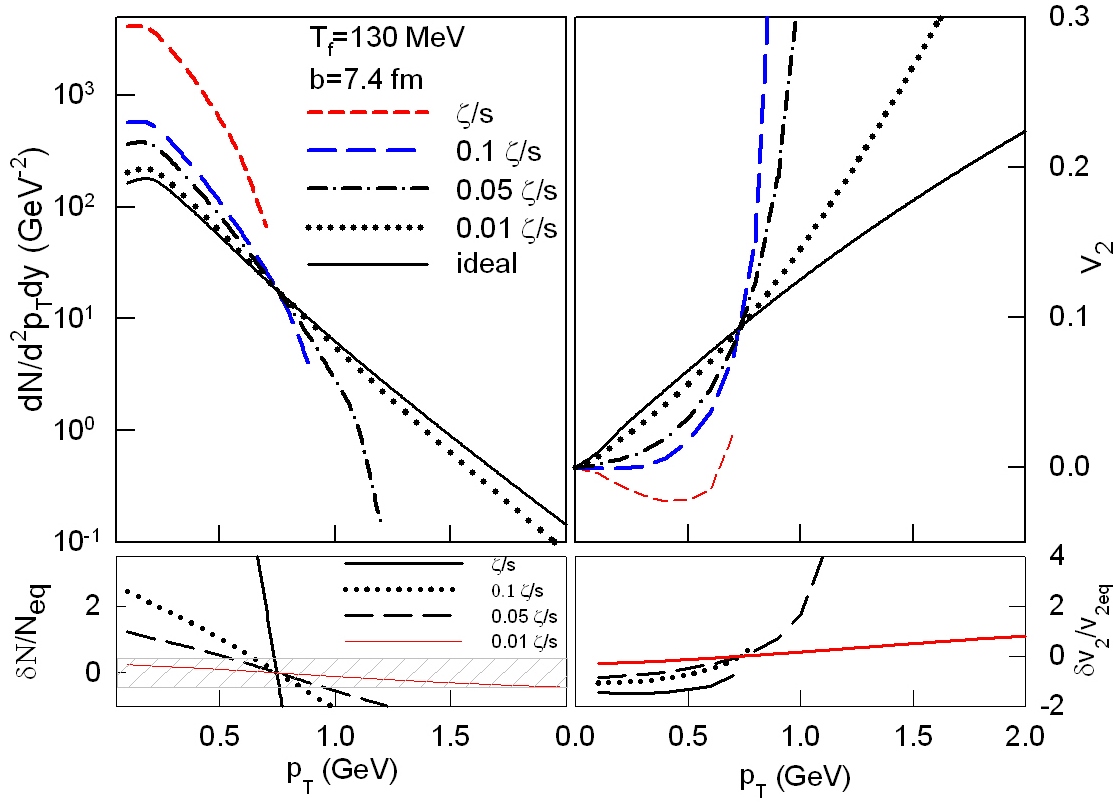}
\caption{(Color online) The invariant yield of $\pi^{-}$ as a function of $p_{T}$ for ideal
and bulk viscous evolution with four different $\zeta/s$ values are shown in the upper
left plot. The lower left plot shows the relative correction to the $\pi^{-}$
yield as a function of $p_{T}$ for four different $\zeta/s$. See text for details.
The right side plot
is same as the left but for $v_{2}$.\label{fig4}}
\end{figure}

We have employed Grad's fourteen-moment method for calculating the dissipative
correction to the freezeout distribution function as described in~\cite{journal-7}.
The details of the implementation of this method
to our viscous code~"`AZHYDRO-KOLKATA"' can be found in~\cite{journal-8,journal-9}. 
The top left panel of figure~\ref{fig4}
shows the $p_{T}$ spectra of pions for ideal(black solid line) and for four different values
of $\zeta/s$. The corresponding relative correction to the $p_{T}$ spectra is shown
in the bottom left panel.
The $v_{2}$ of pion and the relative correction 
is shown in the right panel of figure~\ref{fig4}.
Freeze-out correction in Grad's moment method is obtained under the assumption that 
the non-equilibrium correction to the distribution function is small than the equilibrium 
distribution function. It is then implied that the relative correction $\delta N/N_{eq}$ 
is small for Grad's method to be applicable. 
The shaded band in the bottom left panel corresponds to the relative
correction of 50\%. If we consider here a correction of
magnitude greater than 50\% indicates the breakdown of
the the freezeout correction procedure then our study shows that the Grad's method
will be applicable if the $\zeta/s$ has value less than 0.01 times the present form
considered here.

\end{document}